\documentstyle[epsf]{l-aa}

\def\Teff{effective temperature} 
 
\def\fq{frequency}
\def\fqs{frequencies}
\def\acfq{acoustic cut-off frequency}
\def\atm{atmosphere}
\def\atms{atmospheres}
\def\md{model}
\def\mds{models}
\def\Ku{Kurucz} 
\def\teff{$T_{{\rm eff}}$}
\def\lgg{$\log g$}
\def\roext{$\rho_{{\rm ext}}$}
\def\tauext{$\tau_{{\rm ext}}$}
\def\i{\,{\sc i}} 
\def\nuc{$\nu_{\rm c}$}
\def\nuac{$\nu_{\rm M}$}
\def\nuo{$\nu_{\rm c}^{\rm (1)}$}
\def\nuoM{$\nu_{\rm M}^{\rm (1)}$}
\def\nutw{$\nu_{\rm c}^{\rm (2)}$}
\def\nutwM{$\nu_{\rm M}^{\rm (2)}$}
\def\nuth{$\nu_{\rm c}^{\rm (3)}$}
\def\nuthM{$\nu_{\rm M}^{\rm (3)}$}

\begin{document} 

\thesaurus{06(08.01.3, 08.03.2, 08.09.2, 08.15.1, 08.22.3)}

\title{	The acoustic cut-off frequency of 
	roAp stars\thanks{Based on {\sc hipparcos} data.}} 

\author{N.\,Audard\thanks{Lise Meitner fellow of the
{Fonds zur F\"orderung der wissenschaftlichen Forschung}}$^,$\inst{1,2,3}, 
   F.\,Kupka\inst{1}, P.\,Morel\inst{2},  J. Provost\inst{2},  
   W.\,W.\,Weiss\inst{1}}

\offprints{N. Audard, Cambridge}

\institute{Institute for Astronomy, University of Vienna, 
           T\"urkenschanzstra\ss e 17, 
        A-1180  Vienna, Austria \\ \hspace*{6pt} 
(last\_name@galileo.ast.univie.ac.at) 
        \and  D\'epartement Cassini, UMR 6529, 
                 Observatoire de la C\^ote d'Azur, BP 4229, F-06304 
		 Nice cedex 4, 
            France \\ \hspace*{8pt}(last\_name@obs-nice.fr)
        \and  Institute of Astronomy, University of  Cambridge, Madingley Road, 
            Cambridge CB3 OHA, England  
             \\ \hspace*{8pt}(last\_name@ast.cam.ac.uk)}
         
\date{Received; submitted}

\maketitle

\markboth{N.\,Audard et al.: 
	The acoustic cut-off frequency of Ap stars}
{N.\,Audard et al.: Effects of atmosphere modelling on the acoustic 
cut-off frequency of Ap stars}

\begin{abstract}
Some of the  rapidly oscillating (roAp) stars, have \fqs\ which are larger
than the \acfq\ determined from stellar models with atmospheres based on an
Eddington or Hopf law. 

As the cut-off frequency depends on the $T(\tau )$ relation in the 
atmosphere, we have computed \mds\ and adiabatic \fqs\ for 
pulsating Ap stars with $T(\tau)$ laws based on Kurucz 
model atmospheres and on the Hopf's purely radiative relation. 

We compare the values of the cut-off \fq\ derived from expressions 
for the potential from Vorontsov \& Zarkhov (1989), from 
Gough (1986), and from the approximation of an isothermal atmosphere. 
These models predict a different reflexion efficiency for waves 
and {hence marginally} different values for the 
cut-off \fq. The fre\-quen\-cy-de\-pen\-dent treat\-ment  
of ra\-dia\-ti\-ve trans\-fer as well as an improved calculation 
of the radiative pressure in Kurucz model atmospheres  
increase the theoretical \acfq\ for roAp stars by about 200\,$\mu$Hz, 
which is closer to the observations. 

Since evolutionary effects significantly influence the \acfq\ we
restrict the comparison of our computations with observations to those two
`pathological' roAp stars for which more reliable astrophysical 
parameters are available, HD\,24712 and $\alpha$\,Cir, and
comment briefly on a third one, HD\,134214.
For $\alpha$\,Cir we find models with Kurucz atmospheres which have 
indeed a cut-off frequency 
beyond the largest observed frequency  and which are well within the \teff\ --
$L/L_{\sun}$ error box. For HD\,24712 only models which 
are hotter by about 100\,K and less luminous by nearly 10\% than what is 
actually the most probable value would have an \acfq\ large enough. 
HD\,134214 fits our models best, however,
the error box for \teff\ -- $L/L_{\sun}$ is the largest of all three
stars.

One may thus speculate that the old controversy about a 
mismatch between observed largest frequencies
and theoretical cut-off frequencies of roAp star models is resolved.
However, the model atmospheres have to be refined by investigating NLTE 
effects, among others, and the observational errors for the astrophysical 
fundamental parameters have to be reduced further before a definite 
conclusion can be drawn.
For the latter, asteroseismology can provide an important 
improvement by determining  the frequency splitting  
$\nu_0 = (2~{\int_0^R(dr/c)})^{-1}$ which is sensitive to the 
evolutionary status of pulsating stars.

\end{abstract}

     \section{Introduction}

For 5 out of 28  known rapidly oscillating magnetic chemically peculiar  
(CP2, Preston 1974) stars, the so-called roAp stars, the  largest 
observed frequency exceeds the theoretical \acfq ,
which is determined by the outermost stellar regions. 

Waves with frequencies larger than the cut-off frequency are not well 
reflected  
towards the resonant cavity and decay in the \atm\ with a decreasing 
amplitude.  
It has been argued by Shibahashi and Saio (1985) that the  
cut-off \fq\ is largely influenced by the $T(\tau)$ relation which
requires a careful modelling of these layers.  
Extensive studies of the external layers and of the cut-off \fq\ have 
been carried out for the Sun (e.g., Gough 1986, Balmforth 
\& Gough 1990). Investigations of other types of stars have been neglected 
until
the discovery of roAp stars.

Frequently, atmospheres in stellar \mds\ are based on an
Eddington or Hopf law (e.g. Mihalas 1978, called hereafter 
standard models). In such a case
the $T(\tau)$ relation is given by 
$$T^4 = (3/4)\cdot T_{{\rm eff}}^4 (\tau + q(\tau)),$$ 
where $T$ is the temperature, \teff\ the effective 
temperature, $\tau$ the Rosseland optical depth 
and $q(\tau) \equiv 2/3 $ for the Eddington law.   
These $T(\tau)$ relations suffer from two major approximations:    
convection is not included and radiative transfer is considered 
to be \fq\ independent (i.e. the grey case  is assumed). 
CP2 stars with masses between 1.6 to 2.2\,$M_{\odot}$ have two 
very thin convection zones below the surface. 
The outer zone extends into the atmosphere and it is therefore
necessary to include convection in the model atmospheres.      

The  {\sc atlas}9-code developed by Kurucz (1991)  
takes into account the effect of \fq -dependent opacities from both 
continuum sources and from  about $58$ million lines, 
hence including blanketing effects. 
It also explicitly accounts for convection and radiative pressure.  

In this paper we describe our computations of stellar models where 
atmospheres have been implemented which were derived from the Hopf 
law as well 
as from the more consistent \Ku\ \md\ \atms. 

	\section{Stellar models} 	\label{Smodels}

In stellar structure calculations one separates the inner part, 
where the diffusion ap\-pro\-xi\-ma\-tion for radiative transfer 
is valid, from an outer part, where the atmosphere is interpolated 
using a $T(\tau, T_{\rm eff}, g)$ law, with the gravity $g$.  
When convection is included in a \md\ \atm, the transition 
between the inner part of a stellar model and the atmosphere must be 
located at 
an optical depth $\tau = \tau_{\rm b} \ga 10$, where the diffusion  
approximation is still valid (Morel et al. 1994). 

Effects from a magnetic field are neglected in our stellar models. To discuss
the influence of convection and opacities on a depth dependence of
the critical \fq\ we compare our results obtained with 
advanced \md\ \atm s (based on {\sc atlas}9) with a much simpler grey 
atmosphere
(Hopf) which is a very limited description of a stellar atmosphere,
but which is frequently assumed for stellar models.

	\subsection{Atmosphere}			\label{amod}

In order to describe the \atm\ we used the 
LTE Kurucz {\sc atlas}9 code without the ``overshooting option''
(Castelli 1996) to calculate an interpolation table for $T(\tau, 
T_{{\rm eff}}, g)$.  
        \begin{table}
	\begin{center}
	\begin{small}
	\caption{Fundamental astrophysical parameters for models of 
                 1.8\,$M_{\odot}$ at different evolutionary stages, 
                 with an \atm\ derived from \Ku\ {\sc atlas}9 \md\ \atm s.} 
							\label{Tevol}
	\begin{tabular}{ccccc} 
\hline
age & $\log(T_{{\rm eff}})$ & $\log(L/L_{\odot})$ & 
$R / R_{\odot}$ & $\log g$ \\
($10^6$ yrs) &        &   \\
\hline
40   & 3.9209 & 1.0162 & 1.550 & 4.312\\
100 & 3.9198 & 1.0264 & 1.574 & 4.299 \\
225 &  3.9171 & 1.0426 & 1.624 & 4.272 \\
325 & 3.9140 & 1.0561 & 1.673 & 4.246 \\
400 & 3.9113 & 1.0645 &  1.711 & 4.227 \\
500 & 3.9067 & 1.0772 & 1.773 & 4.196 \\
\hline
	\end{tabular}
	\end{small}
	\end{center}
	\end{table} 
Model \atms\ with solar 
composition were computed for $\log g = 4.2$ and for \teff\ ranging from 
7400 to 10000\,K, with steps of 100\,K, and no
additional contribution to line opacity by microturbulence 
has been assumed. 
The gravity varies in our 1.8\,$M_{\odot}$ model (Tab.\,1)
from $\log g = 4.313$ at the ZAMS to $\log g = 4.196$ at an age of 500\,Myr. 
However, the approximation of a constant \lgg\ ({\em only for the 
atmosphere!}) 
in evolutionary calculations is justified as the $T(\tau, T_{{\rm eff}}, 
\log g$) law is rather insensitive to 
small variations of \lgg .
Gravity effects on the cut-off frequency along evolutionary tracks 
have been studied by, e.g., Shibahashi (1991) and shall be discussed 
in subsection 3.4.  

Generally it is adequate to compute
\Ku\ \mds\ with 72 layers, but the numerical accuracy    
is insufficient when derivatives are important, such as the temperature 
gradient. 
This problem was also mentioned by D. Katz and C. van't Veer 
(priv. comm.) and we have therefore computed all our \atm\ \mds\ with 288 
layers.

For stellar models with a simpler \atm\ based on the Hopf law where convection 
is not treated it is sufficient to restrict the computations for the \atm\ 
down to $\tau_{\rm b} = 2$. 
The values of $q(\tau)$ at each optical depth are interpolated from the values 
given in Table (3.2) of Mihalas  (1978). 

The outer boundary of solar \md\ \atm s is usually set to the bottom of the 
chromosphere, at \tauext\ $ \sim 10^{-4}$, where the 
minimum of temperature occurs for the Sun. Since it is not known whether CP2 
stars have a 
corona or not, we have tentatively defined the outer boundary of our 
atmospheres 
at 
\tauext\ $ = 10^{-6}$ and will justify this choice later. 
The model atmospheres, however, were always computed up to 
$ \tau = 10^{-6.875}$, as in Kurucz (1993) model grids.  
The density at the stellar model boundary was fixed to \roext\ $ = 1.895\cdot 
10^{-11} {\rm g\cdot cm^{-3}}$, 
which is the value for the Kurucz \md\ \atm\ with $T_{{\rm eff}} = 8080$\,K at 
the given
optical depth. 

	\subsection{Internal structure}		\label{inmod}

We have computed representative \mds\  for CP2 stars of 1.8\,$M_{\odot}$ with  
the CESAM code  (Morel 1993 and 1997).  
These models have about 1600 mesh points and include a ZAMS sequence. 
The EFF equation of state (Eggleton et al. 1973) was used, an initial 
hydrogen content $X = 0.7$, and a heavy-element abundance $Z$ = 0.02. 
The most recent OPAL95 opacities (Iglesias \& Rogers 1996) 
with a bi-rational spline interpolation 
(Houdek \& Rogl 1996) were incorporated. 
Convection is described by the standard mixing-length theory  
with a mixing length $\lambda = \alpha H_{\rm p}$  
(where $\alpha \equiv 1.4$ is the mixing-length parameter and $H_{\rm p}$  
is the pressure scale height). 
Because the two external convection zones are very thin, 
the general properties of the models are insensitive to the 
value of $\alpha$ (Gough \& Novotny 1993, Audard \& Provost 1994).  

We shall call a ``Hopf model'' a full stellar model constructed with an 
\atm\ derived from Hopf's law, and a ``Kurucz model'' a model with
a Kurucz model \atm\ implemented. 

The  evolution of a 1.8\,$M_{\odot}$ \Ku\ \md\ is summarized 
%in Tab.\,\ref{Tevol}, and the main characteristics of Hopf and Kurucz models 
in Tab.\,1, and the main characteristics of Hopf and Kurucz models
%with an age of $500\cdot 10^6$ years are given in Tab.\,\ref{Tdiff}.  
with an age of $500\cdot 10^6$ years are given in Tab.\,2. 
As expected, they have similar effective temperatures of about 8066 K, and 
their radii and therefore luminosities are also very similar. 
The positions of the convection zones, which move very little during the 
evolution, are also given. 
	\begin{table}
	\begin{center}
	\begin{small}
	\caption{Characteristics of our 1.8\,M$_{\odot}$ models
		with an age of $500 \cdot 10^6$ yrs. 
		The Hopf  model   has an \atm\ derived from 
		the Hopf law, while the \atm\ of the \Ku\ model is derived 
		from \Ku 's \md\ \atms. 
		The first lines give the effective temperature, luminosity 
		and radius in solar units, and the characteristic 
		\fq\ $\Omega_g =  (G M /R_{\star}^3)^{1/2} $. 
		The second lines indicate the positions of the two 
		outer convection zones in units of the stellar radius ${\rm 
		R}_{\star}$. 
		The deeper one is located between the radii $r_{c1}$ and  
		$r_{c2}$, 
		the ou\-ter\-most con\-vec\-tion zone lies bet\-ween $r_{c3}$ 
		and  $r_{c4}$. 
		The positions of these limits are 
		expressed in logarithm of the pressure in the third lines.  }
								\label{Tdiff}
	\begin{tabular}{rcccc} 
\hline
 model & $T_{{\rm eff}}$ (K) & $L / L_{\odot}$ & $R / R_{\odot}$ & 
$\Omega_g (10^{-6} {\rm rad.s}^{-1}$)\\ 
        &   $r_{c1} / {\rm R}_{\star}$ & $r_{c2} / {\rm R}_{\star}$ & 
          $r_{c3} / {\rm R}_{\star}$ & $r_{c4} / {\rm R}_{\star}$ \\[-3mm] \\
          & $\log P_1$  & $\log P_2$ & $\log P_3$ & $\log P_4$  \\
\hline
\\
   Hopf  &  8067  &  11.94  &  1.772 &   356.9  \\
       &  0.9915 &  0.9946  & 0.9988  &  1.000 \\
       &     5.96    &  5.53   &    4.595 & 3.97 \\
\\
 Kurucz &  8066  &  11.95  &  1.773  &  356.6 \\
       &  0.9914 &  0.9945  & 0.9988   & 0.9999  \\
         &     5.96    &  5.53   &  4.57 & 3.93 \\
\hline
	\end{tabular}
	\end{small}
	\end{center}
        \end{table}
  
	\section{Critical and cut-off frequencies}

In this section we comment on the definitions of the critical
(\nuc ) and cut-off  (\nuac ) frequencies, report on
our calculations of these quantities and compare them with observations. 

	\subsection{Theoretical background}		\label{Sfrequ}

Our oscillation code integrates the $4^{th}$ order
system of equations governing the stellar nonradial
adiabatic oscillations and takes into account 
the gravitational potential perturbations (Unno et al. 1989).  
These equations are obtained by perturbing 
the equations of momentum and continuity.  
They do not include the effects of a magnetic field. 

Because the light received from stars other than the Sun 
is integrated over the visible hemisphere,  
essentially only modes of low degree $\ell$ can be observed.   
We have therefore restricted our calculations to $\ell$ = 0 to 3, and
we consider high radial orders $n$, which justifies 
a Richardson extrapolation between \fqs\ derived from \mds\ 
with about 1600 and 800 points (Shibahashi \& Osaki 1981). 

To calculate the critical and the cut-off frequencies 
we make the following assumptions. 
For low-degree modes, the displacement is essentially vertical,
so that the horizontal component can be neglected.
Consequently, we consider only radial modes and,    
because we investigate modes of high radial orders, 
we adopt the 
Cowling approximation, i.e. neglect the perturbation of 
the gravitational potential. 

The equation of motion for adiabatic oscillations can be written 
(Unno et al, 1989) as: 
$$ \Phi^{\prime \prime} + {\omega^2 - V^2 \over c^2} \Phi = 0, \eqno(1) $$
where $\Phi=\rho^{1/2} c^2 {\rm div}\xi$, and $\xi$ is the  
fluid displacement, $c=({\Gamma_1 P / \rho})^{1/2}$ is the 
sound speed, 
$\omega=2 \pi \nu$ is the angular velocity (with the pulsation frequency 
$\nu$),  
$V$ is the acoustic potential, and the derivative is with respect to the 
radius. 
Eq. 1 shows that acoustic waves are
reflected towards the interior and are well trapped, if $\omega^2$ is 
smaller than $V^2$, whereas if 
$\omega^2$ is larger, the mode propagates into the \atm\ dissipating
mechanical energy  which decreases the mode amplitude.  
The {\em critical \fq,} $\nu_c$, is defined as the \fq\ above which modes 
propagate outwards. The {\em cut-off \fq,} \nuac,  is the maximum value 
of the critical \fq\ encountered in the outermost stellar layers. 

According to Vorontsov \& Zarkhov (1989), the potential  
for radial modes can be written as: 
$$ V_1^2 = N^2 - {c \over 2} {{\rm d} \over {\rm d}r}
{\left[c\left({2 \over r} + {N^2 \over g} - {g \over c^2}
- {1 \over 2 c^2} {{\rm d}c^2 \over {\rm d}r}\right)\right]} +$$
$${c^2 \over 4} \left({2 \over r} + {N^2 \over g} - {g \over c^2}
- {1 \over 2 c^2} {{\rm d}c^2 \over {\rm d}r}\right)^2, \eqno (2) $$
where $N$ is the Brunt-V\"ais\"al\"a frequency.  
Another formulation is proposed by Gough (1986), 
assuming the characteristic length of the
eigenfunctions $({\rm d}\ln (\delta r / r) / {\rm d}r)^{-1}$
being short compared to $r$, so that the problem corresponds to a
plane-parallel layer in constant gravity.   
Under these conditions, the potential reduces to:  
$$ V_2^2 = \omega_{\rm c}^2, 
\eqno (3)$$ 
where $\omega_{\rm c}^2=c^2 / ({4 H_{{\rm \rho}}^2}) 
\left(1 - 2 {{\rm d} H_{{\rm \rho}} / {\rm d}r}\right)$ 
is the critical angular velocity and  
$H_{{\rm \rho}}=-({\rm d}\ln \rho / {\rm d}r)^{-1}$ is the
density scale height.  
Vorontsov \& Zarkhov (1989) use the acoustical depth  as the dependent 
variable, while Gough (1986) uses the radius. 
The approximation of an isothermal \atm\ is also often adopted for 
calculating \nuc, and $H_{{\rm \rho}} = H_{\rm p}$ is then constant.
In this case Eq. (2)  and (3) reduce to: 
$$V_3^2 = \omega_{\rm c}^2 =  {c^2 \over {4 H_{{\rm \rho}}^2}}.  \eqno(4)$$

Rapid density variations in the medium, such as occurring 
at the transition between the bottom of the radiative zone 
and the top of the convective zone, lead to an abrupt rise in  
the potential which constitutes a reflecting barrier to waves. 
But to some extent, waves penetrate and
may be even tunnelling outwards (see e.g., Christensen-Dalsgaard and Frandsen 
1983,
Balmforth \& Gough 1990).
The thickness of the potential barrier, which determines 
the reflexion efficiency, is a function of the outer boundary conditions 
and of 
the potential $V$. We shall see in the next section that the different 
expressions 
for $V$ predict a different 
thickness of this barrier and therefore a different efficiency of the wave 
reflexion. 
In the solar case, a potential barrier occurs at $\tau \sim 10^{-4}$, because 
the temperature rises above these layers and the potential behaves as 
$T^{-1/2}$.
This temperature minimum is related to the boundary between the solar 
photosphere and chromosphere.  

As there is no clear observational evidence for a corona around
CP2 stars, the existence of a temperature minimum and therefore of a 
maximum of 
the potential is not known either (Balmforth \& Gough 1990).    
Note however, that a temperature minimum could be caused also by other
 phenomena
than an outside corona.  
The concept of the \acfq\ as the largest critical \fq, however, requires a 
decrease of the critical frequency after a local maximum is reached in the 
same 
atmospheric layer where a temperature minimum is observed. 
First evidence of a corona in Ap stars has perhaps been recently 
discovered by Simon \& Landsman (1997).

In the next section we shall speculate on the \acfq, provided that a
temperature minimum at an optical depth of either $\tau_{{\rm ext}}=10^{-4}$ 
(as for the Sun) or $10^{-6}$ exists.
The latter depth was chosen, because layers at $\tau = 10^{-6}$ are optically 
transparent for {\em all} 
wavelenghts (Castelli et al. 1997). Higher layers are even more 
sensitive to NLTE effects which are not included in our models, 
while deeper layers become opaque in various wavelength ranges. 
Thus $10^{-6}$ is a reasonable compromise. 

We denote \nuo, \nutw, and \nuth\  as the critical \fqs\ $V / 2\pi$ 
derived from the equations (2), (3) and (4), respectively, and
similarly  \nuoM, \nutwM, and \nuthM\ as the cut-off \fqs, i.e the maximum 
values 
of the critical \fqs\ encountered in the upper atmosphere.

       \subsection{Kurucz versus Hopf models} \label{Sresults}

Figure\,1 gives the temperature profile as a function of  
the logarithm of the optical depth for stellar models of 1.8 $M_{\odot}$ 
with  an atmosphere based on the Hopf law and on \Ku\ \mds, as well as the 
specific Kurucz {\sc atlas}9 \md\ \atm\ for \teff\,=\,8080\,K.  
We see that the \Ku\ \md\ reproduces very well the 
specific \Ku\ \md\ \atm\ of the same \teff. 
In the \atm, the \fq -dependent treatment of radiative transfer
increases the local temperature at a given radius, except  
in the very outer layers where the temperature of the \Ku\ \md\ decreases 
below the limiting value of the Hopf law (Fig. 1b).

The depth dependence of the temperature  
has a direct consequence  on the gradient, the ionization of H and 
He\i, on $\Gamma_1$, and on $c$.  It directly 
affects therefore the critical \fq\ (Eq. 1).  
Fig.\,2  shows the temperature gradient  
$\nabla = {\rm d}\log T / {\rm d}\log P$ and 
the adiabatic temperature gradient $\nabla_{{\rm ad}}$  
as a function  of the logarithm
of pressure, for the same models as in Fig.\,1  
(only the outermost convection zone is 
shown; the deeper  one occurs at $r/R \sim 0.995$, i.e.
at ${\rm log}P \sim 5.5$, see Tab.\,2). 
The superadiabatic temperature gradient  exhibits two close peaks, 
which are related to the transitions between radiation and 
convection and back to radiation at the borders of the thin
convection zone.
For $\tau \geq 0.03$, the local temperature of the Kurucz model 
is larger than that of the Hopf model (Fig.\,1) 
and the peaks of the temperature gradient of the Kurucz \md\ are shifted 
therefore towards the surface compared to the Hopf \md. Similarly,
the ionization zones of H and He\i\ and the peaks of the adiabatic 
exponent $\Gamma_1$ are also closer to the surface 
(Fig.\,3). 

       \begin{figure}
             \epsfxsize = 9cm
             \epsfysize =   9cm
%               \epsffile{tau_MLT_all_fig_3.eps}  
               \caption{Temperature profile of  models with
                 an \atm\ derived from the Hopf law (dotted) and from
                  \Ku\ \md\ \atms\ (dashed).  Both have an effective 
                  temperature of about 8080 K. In the upper panel {\bf (a)}, 
                  the solid line corresponds to the specific Kurucz 
                  model atmosphere with \teff\,=\,8080\,K and $\log 
		  g\,=\,4.2$, 
                 and, as expected, coincide  with the dashed line.
                 The lower panel {\bf (b)} shows the outermost layers.}
              	\label{tprofile}
        \end{figure}

        \begin{figure}
          \epsfxsize = 9cm
          \epsfysize = 4cm
%                \epsffile{grad_hopf_atmo_fig.eps}
          \caption{Temperature gradient $\nabla = {\rm d}\log T / 
	  {\rm d}\log P$ 
                       for the models with an \atm\ derived from Hopf law 
(dotted) and 
                       from \Ku\ \md\ \atms\ (dashed). Both have an effective 
                       temperature of about 8080\,K.  The adiabatic gradient 
                       $\nabla_{{\rm ad}}$ is also presented. The solid line 
                       corresponds to the specific  Kurucz model atmosphere   
                       with \teff\,=\,8080\,K and $\log g=4.2$. }
               							\label{tgrad}
        \end{figure}

At $\tau\,\leq\,0.03,\ (\log P\,\leq\,3.9)$, the temperature gradient is 
radiative (Fig.\,2) and the temperature in the outermost layers 
of the \Ku\ \md\ is smaller by about 1220\,K than that of the Hopf \md\  
(Fig.\,1b). On the other hand, the adiabatic gradient and 
$\Gamma_1$ are larger for the \Ku\ \md\ (Fig.\,3). 
While $\Gamma_1$ reaches a maximum at $\log P \sim 3.7$ for the Hopf \md\ and 
then decreases monotonically, it slightly increases till $\log P \sim  2$ for 
the  
\Ku\ \md\ before decreasing. This behaviour has a direct influence on the 
\acfq, 
as will be shown below.  
$\Gamma_1$ decreases below  4/3 for the Hopf model and tends to 1. The 
Eddington law has the same result and furthermore a similar tendency is found 
(G.\,Houdek 1997, private communication) for an atmosphere derived from the 
semi-empirical solar model `C' from Vernazza et al. (1981). The tendency of 
$\Gamma_1$ to 
decrease below 4/3 reveals the failure of $T(\tau)$ relations based on the Hopf 
law, on Vernazza et al.'s model, or similar, to describe properly the upper 
layers of stars which are more massive than the Sun. 

      \begin{figure}
	\epsfxsize = 9cm
	\epsfysize = 4cm
%               \epsffile{v_hopf_MLT_fig.eps}
	    \caption{Adiabatic exponent 
               $\Gamma_1 = (\partial{\rm ln}P / \partial 
               {\rm ln}\rho)_{{\rm ad}}$ for stellar models with an \atm\ 
		derived 
                  from the Hopf law (dotted), 
                  and from Kurucz models (dashed).} 
            \label{adiabatic}
          \end{figure}

The potential barrier formed by $V_2$ (Eq. 3) 
at the transition between the radiative and the convective zones at 
$\log \tau$ between  1.5 and -1 is larger than for $V_1$ (Eq. 2,   
Fig.\,4). The consequences of an efficiently reflecting 
boundary (represented by $V_2$) and a running wave boundary (approximated by 
$V_1$) 
have been investigated in the solar case by Gabriel (1992) and result in 
different
mode amplitudes. In the upper layers 
the agreement between the three expressions for the potential 
is better for the Kurucz model (Fig. 4b) than for the Hopf \md\,(Fig. 4a).  
The discrepancies 
at $\tau \le 10^{-4}$ are probably due to NLTE and nonadiabatic effects
where the main assumptions for the formal expressions of the potentials break 
down. 

	\begin{figure}
	\epsfxsize = 9cm
	\epsfysize = 9cm
%               \epsffile{pot_fig.eps}
	\caption{{\bf a:} (top)  Logarithm of the potentials $V_1$, 
		$V_2$ and $V_3$ (Eq.\,2, 3 and 4, full, dotted and dashed 
		lines,
		respectively), in the upper layers of the Hopf model;  
		{\bf b:} (bottom) same for the Kurucz \md.}
                						
\label{potentials}
        \end{figure}

	\begin{table}
	\begin{center}
	\begin{small}
        \caption{Acoustic cut-off frequencies 
		\nuoM, \nutwM\ and \nuthM\ in $\mu$Hz (see Eqs.\,2, 3 and 4),  
        	for the Hopf and \Ku\ \mds. The boundary of the model 
		atmospheres  
		is assumed either at \tauext$\,=\,10^{-4}$ or at $10^{-6}$.  
                The three last lines indicate the difference $\Delta$   
                between the respective Kurucz and Hopf models. 
		For \tauext$\,=\,10^{-6}$, we
		also give the cut-off frequencies for models with an 
		artificially 
		suppressed convection in the outer layers (see the next 
      		subsection).}  
                					\label{Tfrequ}
	\begin{tabular}{rccccc} \hline
Model  &          & \tauext: $10^{-4}$ & $10^{-6}$ &  (no convection)\\
\hline
\\
Hopf   &  \nuoM   & 2139 & 2324 & (2061) \\
       &  \nutwM  & 2159 & 2354 & (2345) \\
       &  \nuthM  & 2231 & 2232 & (2224) \\
\\
Kurucz &  \nuoM   & 2453 & 2528 & (2568) \\
       &  \nutwM  & 2495 & 2743 & (2748) \\
       &  \nuthM  & 2463 & 2465 & (2470) \\
\\
\\
  & $\Delta \nu_{{\rm M}}^{\rm 1}$ &  314 &  204     \\
  & $\Delta \nu_{{\rm M}}^{\rm 2}$ &  336 &  389     \\
  & $\Delta \nu_{{\rm M}}^{\rm 3}$ &  232 &  233     \\
\hline
	\end{tabular}
	\end{small}
	\end{center}
        \end{table}

The behaviour of the critical frequencies  \nuo, \nutw, and \nuth\ for 
the Hopf and Kurucz \mds\ are compared in Fig.\,5. In all cases, 
the potential in the upper layers is larger for the Kurucz \md\ 
(at $\tau \le 0.01$). Note that the maximum value of \nuth, the \acfq\ \nuthM, 
does not 
occur at the smallest optical depth, but slightly deeper in the atmosphere. 
 
Table\,3 gives the values of the cut-off \fqs\ \nuoM, \nutwM\ and 
\nuthM\ for the Hopf and Kurucz \mds, computed for the potentials V$_1$, V$_2$, 
and
V$_3$, respectively, and with the assumption that the atmospheric
boundary is located either at \tauext$\,=\,10^{-4}$ or at $10^{-6}$.
The differences of the \fqs\ computed for the Hopf models relative to the 
Kurucz 
models
are substantial and amount to up to 10\% of \nuac.
We must, however, keep in mind that the results derived from the Hopf law 
must be considered  with much caution in optically thin regions,   
i.e. at $\tau \la 10^{-3}$.   

        \begin{figure}
       \epsfxsize = 9cm 
      \epsfysize = 12cm
%       \epsffile{pot_H_K_fig_2.eps}
	\caption{{\bf a:} (top) Critical frequency \nuo\ (Eq.\,2) 
          for the Hopf (dotted) and Kurucz  \mds\  (dashed); 
		{\bf b:} (center) same for \nutw\ (Eq.\,3); 
		{\bf c:} (bottom) same for \nuth\ (Eq.\,4).}
                						\label{critf}
        \end{figure}

We have also compared  the \fqs\ of acoustic modes of low and 
high radial order derived from the Hopf and Kurucz models.  
The latter have \fqs\ larger  by  about  1\,$\mu$Hz
compared to their Hopf counterparts. Although it represents less than 
0.01\% of the \fq, this difference is larger than the 
measurement error of  ground-based observations and 
for the asteroseismic space mission COROT (Catala et al. 1995).  
 
	\subsection{Effects of convection} 

As we have demonstrated in the previous subsection, Kurucz model 
atmospheres, taking convection and \fq-dependent treatment of radiative 
transfer 
into account, considerably improve stellar oscillation models in the upper 
layers. 
To specify which of these effects has the largest influence  on the 
\acfq, we investigate Kurucz \mds\ with a mixing-length parameter 
$\alpha = 0$, hence turning off 
convection,  
and compare them with Hopf models. For the internal structure models  
we have imposed the gradient to be radiative throughout the  
envelope, and the convective core, of course, is not affected.  
Specific Kurucz {\sc atlas}9-\md\ \atms\ were also computed with $\alpha = 0$. 
A comparison should therefore isolate the importance of the radiative transfer
relative to convection.    

In the models without convection ionization of H and He\i\ 
occurs closer to the surface. As a result, the variations of 
$\Gamma_1$ and of the temperature gradient also occur closer to the surface    
(Fig.\,6). Since the outermost stellar layers are radiative 
the physical quantities are unchanged for  models  
without convection (Fig.\,6) compared to models with 
convection (Fig.\,2). Consequently, the cut-off frequencies  
have almost the same values (see Tab.\,3 for the 
case of \tauext\,=\,$10^{-6}$).
        \begin{figure}
                \epsfxsize = 9cm
                \epsfysize = 4cm
%                \epsffile{grad_alpha.eps}
                \caption{Temperature gradient of models with an \atm\ computed 
                        with the Hopf $T(\tau)$ law (dotted), and with Kurucz 
                       \mds~(dashed).   
                        The mixing-length parameter $\alpha$ was set to 0 
			in oder to suppress convection. The solid line  
represents 
			the gradient of the \Ku\ {\sc atlas}9-\md\ \atm\ with 
			\teff\,=\,8080\,K. }
                					\label{tcomp}
        \end{figure}

These results clearly indicate that the major factor influencing the \acfq\ is  
the inclusion of a \fq-dependent treatment of radiative transfer and
a better calculation of the radiative pressure rather than   
the treatment of convection in the atmosphere.
This result was indeed expected since the outermost layers, 
which are relevant for the cut-off \fq, are radiative. 

	\subsection{Effects of evolution}		\label{evol}

The cut-off frequency also depends on the evolution, as was shown for 
example by Shibahashi (1991).  From the ZAMS to the end of the main sequence   
and for models 
of 1.8 M$_{\odot}$ this quantity decreases from 2855  to 1212 $\mu$Hz for the 
Hopf models and from 3116 to 1328 $\mu$Hz for the Kurucz models. 
However,  the relative difference between 
the cut-off frequencies from the Kurucz and Hopf models remains almost 
unchanged 
along the main sequence and is about 8.5\% (see Tab.\,4). 
	\begin{table}
	\begin{center}
	\begin{small}
        \caption{Acoustic cut-off frequencies $\nu_{\rm M}^{(1)}$ (Eq. 2) as a 
                 function of age for 
		 Hopf and Kurucz models with 1.8\,$M_{\sun}$ and Z\,=\,0.02. 
		 The relative difference $\Delta$ is  Kurucz-Hopf. } 
               					\label{Tevol}
	\begin{tabular}{rccc} 
\hline
age (Myr) & Hopf & Kurucz &  $\Delta$ in \% \\
     & ($\mu$\,Hz) & ($\mu$\,Hz) & \\             %  of $\mu {\rm Hz}_{Kurucz}$ 
\\
\hline
 100 & 2855 & 3116 & 8.37 \\
 200 & 2732 & 2985 & 8.48 \\
 300 & 2610 & 2864 & 8.86 \\
 400 & 2481 & 2726 & 8.99 \\
 500 & 2360 & 2581 & 8.56 \\
 600 & 2221 & 2426 & 8.45 \\
 800 & 1914 & 2073 & 7.66 \\
1000 & 1510 & 1650 & 8.49 \\
1100 & 1212 & 1328 & 8.76 \\
\hline
	\end{tabular}
	\end{small}
	\end{center}
        \end{table}
The cut-off frequency also scales as the 
characteristic frequency $\Omega_g = {({\rm G} M / R^3)}^{1/2}$
(or as the frequency spacing, see Shibahashi 1991).   
We obtain similar quantitative results for all masses
typical for roAp stars. 
When investigating a large parameter space in mass, age and metallicity,
it appears to first order to be sufficient to compute 
models with the simple Hopf law and to approximate the effects of a better 
treatment of the atmosphere on the \acfq\ by 
increasing this frequency by about 7\% to 9\%.
 
Since the uncertainty in the age determination can introduce a significant 
error 
in the computed \acfq\ we will focus our discussion in the following
sections on those two roAp stars, HD\,24712 and HD\,128898, for which we
have more reliable mass estimates due to the availability of 
{\sc hipparcos} parallaxes.

	\subsection{Comparison with observations}	\label{Sdiscuss}

For about 5 out of 28 known roAp stars, the largest published \fq\ 
(see Tab.\,5 and references therein) exceeds the expected theoretical 
\acfq\ determined from standard stellar models.
We do not consider here roAp stars for which the highest observed 
frequency probably is a harmonic of their nonlinear oscillation 
(HD\,83368 (Kurtz et al. 1993), 
HD\,101065 (Martinez \& Kurtz 1990), HD\,137949 (Kurtz et al. 1991), 
and HD\,161459 (Martinez et al. 1991)). 

As we have demonstrated in the previous sections, stellar models with a
frequency dependent treatment of the radiative transfer in the atmosphere 
(Kurucz models) do have a higher cut-off frequency than models with a grey
atmosphere (Hopf models) which used to be the baseline for most of the 
investigations in this field. This result is in agreement with speculations
of Shibahashi \& Saio (1985) and of Matthews et al. (1990, 1996)  that a 
steeper than solar temperature gradient would increase the cut-off \fq\
and hence bring theoretical results closer to the observations.  

	\begin{table}
	\caption{List of roAp stars for which the largest published frequency 
		is larger than the cut-off frequency determined from standard 
		stellar models.}
								\label{cut_obs} 
	\vspace*{10pt}
	\begin{tabular}{rccl}
\hline
HD     & Observed                  &  Ampl.       & Comment \\
       &  $\nu _{max}$ ($\mu$Hz)   &  mmag \\
\hline 
\\
  6532 & 2\,402 & 0.77 & Kurtz  et al. (1996) \\
 24712 & 2\,807 & 0.20 & Kurtz et al. (1989) \\
128898 & 2\,566 & 0.12 & Kurtz et al. (1994) \\
134214 & 2\,950 & 3.40 & Kurtz et al. (1991)\\
203932 & 2\,838 & 0.17 & Martinez et al. (1990) \\
\hline
	\end{tabular}
	\end{table}

The photometric and spectroscopic properties of the roAp star {\bf HD\,24712}    
with \teff\,=\,$7250 \pm 150$\,K (Ryabchikova et al. 1997), 
$\log(L/L_{\odot})=0.91 \pm 0.04$ (based on  
$\pi_{\rm HIPPARCOS} = 0\farcs 02041 \pm 0\farcs 00084$, a bolometric 
correction 
of -\,0.085 (Schmidt-Kaler 1982) and 
neglecting interstellar extinction), can be reproduced with a model of 
1.63\,$M_{\odot}$, $Z=0.02$ and an age of about 900\,Myr. 
At \tauext $\,=\,10^{-6}$, the Hopf model gives a cut-off 
\fq\ \nuoM\,=\,2280\,$\mu$Hz whereas the Kurucz model gives 2480\,$\mu$Hz. 
A stellar model with a Kurucz atmosphere hotter by about 100\,K, less
luminous by nearly 10\%, and less evolved by 100\,Myr would have a cut-off 
\fq\ in agreement with the largest observed \fq . However, such a model
is compatible only with the lower left corner of the error box 
(see Fig. 7a). 

We have also computed Hopf and Kurucz \mds\ with appropriate age for 
{\bf HD\,128898} ($\alpha$\,Cir). For the first time, a Kurucz 
\md\ \atm\ was calculated with an opacity distribution function  
specific to the composition of $\alpha$\,Cir (Piskunov \& Kupka 1997).  
Stellar models with 1.93\,$M_{\odot}$, 
$Z\,=\,0.03$ and an age of 400\,Myr, fit the observed 
values (Kupka et al. 1996) of \teff\,=\,$(7900 \pm 200)$\,K 
and $\log(L/L_{\odot}) = 1.11 + 0.01/-0.02$ 
(based on $\pi_{\rm HIPPARCOS} = 0\farcs 06097 \pm 0\farcs 00058$,
a bolometric correction of -\,0.12 (Schmidt-Kaler 1982) and 
neglecting interstellar extinction).  
The cut-off frequencies \nuoM\ for the Hopf and Kurucz models
are 2346  and 2600\,$\mu$Hz, respectively, at \tauext $=10^{-6}$.
The \acfq\ computed for the Kurucz model is compatible with the largest
observed \fq\ (see Table 5) and we can therefore conclude, that no
discrepancy may exist for this roAp star between theoretical and observed 
cut-off \fqs . 

However, the cut-off frequency depends on the model input parameters 
and one must 
therefore account for uncertainties inherent to observations and modelling.
For example, there are problems with the photometric calibration 
of fundamental parameters of CP stars. Models of different mass 
and age can fit the same star in the H-R diagram within the error bars. 
For HD\,24712, e.g., the cut-off frequency varies from 2725\,$\mu$Hz
(Kurucz model with 1.60\,$M_{\odot}$ and 800\,Myr) which is only
80\,$\mu$Hz short of the largest observed frequency, 
to an even lower value of 2294\,$\mu$Hz (1.63\,$M_{\odot}$, 1100\,Myr). 
For $\alpha$ Cir, a model with 1.90\,$M_{\odot}$ and 600\,Myr gives 
\nuoM\,=\,2329\,$\mu$Hz which would be clearly smaller than the largest
observed \fq . 

A similar situation exists for {\bf HD\,134214} which, unfortunately,
has a considerably larger error
box due to the {\sc hipparcos} parallax of only 10.92$\pm$0.89\,marcsec 
and a \teff\ which could be estimated only from photometric indices. 

Evolutionary tracks and lines of constant cut-off frequency for Kurucz 
models are plotted in Fig.\,7a together with the 
observational error boxes. Models with $Z=0.02$ were investigated
for HD\,24712 and HD\,134214, and  models with $Z=0.03$ for 
HD\,128898.  
The same physics as for our previous 1.8\,$M_{\odot}$ models was used 
(see Sec.\,2). Errors on the mass and age determination are about
0.02\,$M_{\odot}$ and 200 Myr for HD\,24712, and 0.02\,$M_{\odot}$ 
and 100 Myr for HD\,128898. 

Since the cut-off frequency of Kurucz models is larger than for
Hopf models of same age, lines corresponding to the Hopf models 
are shifted to higher effective temperature and lower 
luminosity. For $Z=0.02$ the Hopf lines for $\nu_{\rm M}^{(1)}=2300, 2500$ and  
2800\,$\mu$Hz almost coincide with the Kurucz lines  for 
$\nu_{\rm M}^{(1)}=2500, 2800$ and 3000\,$\mu$Hz, respectively, 
and for $Z=0.03$ the Hopf lines for $\nu_{\rm M}^{(1)}=2300$ and 2500\,$\mu$Hz 
coincide with the Kurucz lines for $\nu_{\rm M}^{(1)}=2500$ and 2600\,$\mu$Hz, 
respectively.  

Asteroseismology is a powerful tool for 
determining the evolutionary status of stars via the 
frequency separation 
$$\nu_0 = (2~{\int_0^R(dr/c)})^{-1}  
\sim \nu_{n, \ell} - \nu_{n-1, \ell}$$
(see e.g. Shibahashi 1991, and Kurtz \& Martinez 1993).   
If $\nu_0$ can be measured, a more reliable 
estimate of the evolutionary status can be derived than with 
our classical approach, because no bolometric correction (determined for
chemically `normal' stars) and interstellar 
extinction (with large local differences) have to be considered. 
Fig.\,7b  shows lines of constant frequency spacing 
$\nu_0$ for the same models as for Fig.\,7a. 
The observed value $\nu_0=68\,\mu$Hz for HD 24712 (Kurtz et al. 1989) 
is consistent with our classically determined error box and indicates
an \Teff\ which should be larger by about 100\,K than what was obtained
spectroscopically. There is a serious problem for $\alpha$ Cir, because 
the observed value of $\nu_0$ is $50\,\mu$Hz (Kurtz et al. 1994) 
which cannot be reconciled with the spectroscopically determined \Teff\ and/or
the luminosity.  % derived via the {\sc hipparcos} parallax. 
One has to stress, 
however, that the amplitudes for the overtone oscillations relative to the 
mode with the largest amplitude are very small and of the order of only a few 
0.1\,mmag\,! 

       \begin{figure}
                \epsfxsize = 9cm
                \epsfysize = 12cm
%                \epsffile{hr_fig_acfq_nu0_2.eps}
                \caption{HR diagramme for stars with 1.58 $M_{\odot}$ 
                to 1.69 $M_{\odot}$ for $Z=0.02$ 
                (age up to 1000 Myr), and with 1.90 and 1.93 
                $M_{\odot}$ for $Z=0.03$ (age up to 700 Myr) 
                (dashed lines).  The roAp
                stars HD\,24712, HD\,134214 and HD\,128898 are indicated 
                by circles and error boxes. 
                Full lines are  
                lines of  constant cut-off frequency  
                $\nu_{\rm M}^{(1)}$  for the Kurucz models, for
                2300, 2500, 2800 and 3000\,$\mu$Hz for $Z=0.02$, 
                and for 2300, 2500 and 2600\,$\mu$Hz for $Z=0.03$  
                ({\bf a}), \newline
		and lines of constant frequency 
                spacing $\nu_0 = (2~{\int_0^R(dr/c)})^{-1}$ 
                for the same models, from 80  
                to 65\,$\mu$Hz for  $Z=0.02$, and from 70 to 
                60\,$\mu$Hz for  $Z=0.03$. For HD\,24712
                the frequency splitting $\nu_0$ = 68\,$\mu$Hz (Kurtz et al. 
		1989), 
		and 50\,$\mu$Hz (Kurtz et al. 1994)
		for HD\,128898 ({\bf b}).}
        \end{figure}

We stress also that the frequency difference between models with atmospheres
computed either with the Hopf law or with Kurucz models is comparable to the 
changes which are obtained when introducing physical processes such as 
convective core overshooting (see, e.g., Audard et al. 1995).

	\section{Conclusion}

We  have shown that the major impact for modelling pulsation \fqs\ close 
to the cut-off \fq\ comes from the inclusion of a frequency-dependent 
treatment of radiative transfer and of blanketing effects, as well as 
from a better calculation of the radiative pressure in the \md\ \atm s,  
rather than the inclusion of convection.  
At the main sequence, Kurucz model atmospheres merged with stellar models 
derived from the CESAM code increase the cut-off \fq\ by 
about 8.5\,\% relative to the value derived from the Hopf $T(\tau)$ relation. 

For two roAp stars with probably the best available mass and 
luminosity estimates, HD\,24712 and $\alpha$\,Cir, 
we find models with Kurucz atmospheres and with parameters in agreement with 
the observational error box which have a theoretical cut-off frequency 
larger than the largest observed \fq\ and hence are in agreement with 
observations. One may thus speculate that the old controversy about 
a mismatch between observed largest frequencies
and theoretical cut-off frequencies of roAp star models is resolved.

We have assumed solar abundances for our models except for 
$\alpha$\,Cir, which, however, has an abundance pattern 
that does not significantly change the model atmosphere from  
one with solar abundances. 
HD\,24712, on average, is even less peculiar than $\alpha$\,Cir.
Assuming that Z\,=\,0.03 reflects only a surface composition
for $\alpha$\,Cir and Z\,=\,0.02 would be the better choice for 
a stellar model, one can expect
a smaller mass to fit $\alpha$\,Cir 
and a smaller cut-off frequency as deduced from Fig. 7. 
For other roAp stars with a more peculiar abundance 
pattern, the deviation from model atmospheres with solar abundances or 
models with a scaled heavy element abundance will be larger (Gelbmann 1997). 
Abundant rare-earth elements, through blanketing effects,
could decrease the surface temperature and thus increase the cut-off \fq.
This \fq\ might also be affected by a chemical composition gradient
(Vauclair \& Dolez 1990). Finally, improving CP2 star pulsation models
requires also to account for the magnetic field (Dziembowski \& Goode 1996). 

The low surface temperature in the outer atmospheric regions provides 
a potential well which reflects the acoustic waves  inwards. 
The existence of a maximum value of this potential, 
which defines the cut-off frequency,  requires an increase of the temperature 
in even higher atmospheric layers. 
While such an increase is observed in the solar chromosphere towards the 
corona, 
the first direct evidence of a chromosphere in Ap stars has only been recently 
shown by Simon \& Landsman (1997).  
However, in the very outer atmospheric layers 
nonadiabatic and NLTE effects are probably relevant 
and should be considered in the future. Already at
$\tau \la 10^{-4}$ NLTE effects might lead to an increase of 
temperature (Mihalas 1978) which would decrease the acoustic potential, 
thus defining the upper limit of the critical \fq, and hence the cut-off \fq. 
Unfortunately, NLTE atmospheres with a  treatment of line blanketing 
as sophisticated as in {\sc atlas9}, for at least the main elements 
contributing 
to the opacity in the upper atmospheric regions, are not available in the 
foreseeable future.

        \acknowledgements 
We are indebted to Maurice Gabriel  for his detailed remarks 
which helped improving the manuscript, and to Douglas Gough and G\"unter 
Houdek for their constructive comments. 
We are grateful to Claude van't Veer, Claude M\'egessier and Rodney Medupe  
for stimulating discussions. 
We also thank J\o rgen Christensen-Dalsgaard and Nikolai Piskunov 
for their useful comments. We are also thankful to the referee, Hiromoto 
Shibahashi, for his comments and suggestions. 

This research was done within the working group 
{\em Asteroseismology-AMS}. Computing resources
and financial support for this international collaboration
were provided by the Fonds zur F\"orderung
der wissenschaftlichen  Forschung project {\em S 7303-AST}
(Lise Meitner Fellowship
granted for NA, project {\em M00323-AST}), and Digital
Equipment Corp. (Europe External Research Program,
project {\em STARPULS}). NA is grateful to PPARC  (UK) 
for financial support.


\begin{thebibliography}{}
\bibitem[]{}
Audard, N., Provost, J. 1994, A\&A, 282, 73

\bibitem[]{}
Audard, N., Provost, J., Christensen-Dalsgaard, J.  1995, A\&A,  297, 427

\bibitem[]{}
Balmforth,  N.J., Gough, D.O. 1990, ApJ, 362, 256 

\bibitem[]{}
Castelli, F. 1996, in  {\it Model Atmospheres and Spectrum Synthesis}, 
$5^{th}$ vienna Workshop, Eds. S.J. Adelman, F. Kupka \& W.W. Weiss, 
ASP Conf. Ser., vol. 108,  p. 85

\bibitem[]{}
Castelli, F.,  Gratton, R.G., Kurucz R.L. 1997, A\&A, 318, 841

\bibitem[1995]{}
Catala, C., Mangeney, A., Gauthier, D., Auvergne, M., Baglin, A.,
Goupil, M.J., Michel, E., Zahn, J.P., Magnan, A. Vuillemin, A., Boumier, P.,
Gabriel, A., Lemaire, P., Turck-Chi\`eze, S., Dzitko, H., Mosser, B.,
Bonneau, F.  1995,  GONG'94: {\it ``Helio-and
Asteroseismology From the Earth and Space''},
ASP Conf. Ser. vol. 76, p. 426, Eds. R.K. Ulrich,
E.J. Rhodes, W. D\"appen

\bibitem[]{}
Christensen-Dalsgaard, J., Frandsen, S. 1983, Solar Physics, 82, 165

\bibitem[]{}
Dziembowski, W.A.,  Goode, P.R. 1996, ApJ, 458, 338

\bibitem[]{}
Eggleton, P.P., Faulkner, J., Flannery,  B.P. 1973, A\&A, 23, 325
 
\bibitem[]{}
Gabriel M. 1992, A\&A,  265, 771

\bibitem[]{}
Gelbmann M. 1997, PhD thesis, University of Vienna 

\bibitem[]{}
Gough, D.O. 1986, in {\it Hydrodynamic  and magnetohydrodynamic
problems in the Sun and stars}, Ed. Y. Osaki, University of Tokyo press, 
p. 117 

\bibitem[]{}
Gough, D.O. 1990, in  {\it Progress of seismology of the Sun and stars}, 
Eds. Osaki Y., Shibahashi H., Springer Verlag, p. 283 

\bibitem[]{}
Gough, D.O., Novotny,E. 1993, in  {\it  Inside the Stars},  ASP Conf. Ser., 
Eds.  W.W. Weiss \& A. Baglin, vol. 40, p. 550

\bibitem[]{}
Houdek, G., Rogl, J. 1996, {\it Bull. Astron. Soc. India},  vol. 24, p. 317

\bibitem[]{}
Iglesias, C.A., Rogers, F.J. 1996, ApJ, 464, 943

\bibitem[]{}
Kupka, F.,  Ryabchikova, T.A., Weiss, W.W., Kuschnig, R., Rogl, J., Mathys, G. 
1996, A\&A, 308, 886 

\bibitem[]{}
Kurucz, R. L. 1991, {\it Stellar Models: Beyond  Classical Models}, 
Eds. L. Crivellari, I. Hubeny and D.G. Hummer,  NATO ASI Series, 
Kluwer, Dordrecht 1991

\bibitem[]{}
Kurucz, R. L. 1993, {\it CDROM13: {\sc atlas}9}. SAO, Harvard, Cambridge

\bibitem[]{}
Kurtz, D.W. 1991, MNRAS 249, 468

\bibitem[]{}
Kurtz, D.W., Matthews, J.M., Martinez, P., Seeman, J., Cropper, M., Clemens, 
J.C.,
Kreidl, T.J., Sterken, C., Schneider, H., Weiss, W.W., Kawaler, S.D.,
Kepler, S.O., van der Peet A., Sullivan D.J., Wood H.J., 1989,  MNRAS 
240, 881

\bibitem[]{}
Kurtz, D.W., Kreidl, T.J., O'Donaghue, D., Osiop, D.J., Tripe, P. 1991,
MNRAS 251,152

\bibitem[]{}
Kurtz, D.W.,  Kanaan, A., Martinez, P. 1993, MNRAS 260, 343

\bibitem[]{}
Kurtz, D.W.,  Martinez P. 1993, in  {\it Peculiar versus normal phenomena in 
A-type and 
related stars}, ASP Conf. Ser., Eds. M.M. Dworetsky, F. Castelli, R. 
Faraggiana, 
vol. 44, p. 561

\bibitem[]{}
Kurtz, D.W., Sullivan, D.J., Martinez, P., Tripe, P. 1994, MNRAS, 270, 674

\bibitem[]{}
Kurtz, D.W., Martinez, P., Koen, C., Sullivan, D.J. 1996, MNRAS, 281, 883

\bibitem[]{}
Martinez, P. 1996,  {\it Bull. Astron. Soc. India},  vol. 24, p. 359

\bibitem[]{}
Martinez, P., Kurtz, D.W. 1990,  MNRAS 242, 636

\bibitem[]{}
Martinez, P., Kurtz, D.W., Heller, C.H. 1990, MNRAS 246, 699

\bibitem[]{}
Matthews, J. M., Wehlau, W. H., Walker, G. A. 1990, ApJ, 365, L81

\bibitem[]{}
Martinez, P., Kurtz, D.W. Kauffmann G.M. 1991, MNRAS 250, 666

\bibitem[]{}
Matthews, J. M., Wehlau, W. H., Walker, G. A. 1996, ApJ, 459, 278

\bibitem[]{}
Mihalas, D. 1978, {\it ``Stellar atmospheres''}, Eds W.H. Freeman and Compagny, 
San Fransisco

\bibitem[]{}
Morel, P. 1993, in  {\it Inside the Stars},  ASP Conf. Ser., 
Eds. W.W. Weiss \& A. Baglin, Springer Verlag, vol. 40, p.   445, 

\bibitem[]{}
Morel, P. 1997, A\&AS, in press. Available at 
http://www.obs-nice.fr/morel/CESAM.html

\bibitem[]{}
Morel, P., van't Veer, C., Provost, J., Berthomieu, G., Castelli, F., 
Cayrel, R., Goupil, M.J., Lebreton, Y. 1994, A\&A, 286, 91

\bibitem[]{}
Piskunov, N.E., Kupka, F. 1997, A\&A, in preparation

\bibitem[]{}
Preston, G.W. 1974, ARA\&A 12, 257

\bibitem[]{}
Ryabchikova, T.A., Landstreet, J.D., Gelbmann, M.J., Bolgova, G.T.,
Tsymbal, V.V., Weiss, W.W. 1997, A\&A submitted

\bibitem[]{}
Schmidt-Kaler, Th. 1982, Landolt-B\"ornstein, New Series, VI/2b, Eds. K. 
Schaifers 
\& H.H.Voigt, Springer-Verlag Berlin, p. 452 

\bibitem[]{}
Shibahashi, H. 1991, in {\it Challenges to theories of the structure 
of moderate-mass stars}, Eds. D. Gough \& J. Toomre, 
Lecture Notes in Physics No. 388, Springer-Verlag, p. 393

\bibitem[]{}
Shibahashi, H., Osaki, Y. 1981, PASJ, 33, 713 

\bibitem[]{}
Shibahashi, H., Saio, H. 1985, PASJ, 37, 245

\bibitem[]{}
Simon, T., Landsman W. B. 1997, ApJ, 483, 435

\bibitem[]{}
Unno, W., Osaki, Y., Ando, H., Saio, H., Shibahashi, H. 1989, {\it Nonradial 
Oscillations of Stars}, University of Tokyo press, $2^{nd}$ edition

\bibitem[]{}
Vauclair, S., Dolez, N. 1990, in {\it Progress of Seismology of the Sun and 
stars}, {\it Lecture Notes in Physics}, Eds. Y. Osaki \& H. Shibahashi, 
Springer-Verlag, p. 399

\bibitem[]{}
Vernazza, J.E., Avrett, E.H., Loeser, R, 1981, ApJS, 45, 635 

\bibitem[]{}
Vorontsov, S.V., Zharkov, V.N., 1989, Astro. Sp. Phy. Rev., vol, 7, 
part 1
        \end{thebibliography}
\end{document}